\documentclass[cits]{PoS}
\title{Deflated BiCGStab for linear equations in QCD problems
        \thanks{This work was partially supported by the National Science Foundation, Computational Mathematics Program,
                under grant 0310573.}}

\ShortTitle{Deflated BiCGStab for linear equations in QCD problems}

\author{\speaker{Abdou M. Abdel-Rehim}\\
        Department of Physics, Baylor University, Waco, TX 76798-7316, USA.\\
        E-mail: \email{Abdou$\_$Abdel-Rehim@baylor.edu}}

\author{Ronald B. Morgan\\
        Department of Mathematics, Baylor University, Waco, TX 76798-7328, USA.\\
        E-mail: \email{Ronald$\_$Morgan@baylor.edu}}
        
\author{Walter Wilcox\\
        Department of Physics, Baylor University, Waco, TX 76798-7316, USA.\\
        E-mail: \email{Walter$\_$Wilcox@baylor.edu}}

\abstract{
The large systems of complex linear equations that are generated in QCD problems often have multiple right-hand sides (for multiple sources) and multiple shifts (for multiple masses). Deflated GMRES methods have previously been developed for solving multiple right-hand sides. Eigenvectors are generated during solution of the first right-hand side and used to speed up convergence for the other right-hand sides. Here we discuss deflating non-restarted methods such as BiCGStab. For effective deflation, both left and right eigenvectors are needed.
Fortunately, with the Wilson matrix, left eigenvectors can be derived from the right eigenvectors. We demonstrate for difficult problems with kappa near $\kappa_c$ that deflating eigenvalues can significantly improve BiCGStab. We also will look at improving solution of twisted mass problems with multiple shifts. Projecting over previous solutions is an easy way to reduce the work needed.
}

\FullConference{The XXV International Symposium on Lattice Field Theory\\
                 July 30 - August 4 2007\\
                 Regensburg, Germany}
\begin{document}

\section{Introduction}
Current lattice QCD simulations attempt to reach physical up and down quark masses. 
In this regime, standard linear solvers used in quark propagator calculations converge
very slowly. Roughly speaking, the rate of convergence is proportional to the square root
of the ratio of the smallest eigenvalue to the largest eigenvalue of the Dirac matrix. 
A remedy of this problem is to deflate some of the eigenvectors corresponding to the 
smallest eigenvalues \cite{lat07-Wilcox-plenary}. For restarted GMRES, this was done by
augmenting  the Krylov subspace with approximate eigenvectors with small eigenvalues. The
resulting algorithm is called GMRES with deflated restarting or GMRES-DR \cite{Morgan-Wilcox-gdr}.
One advantage of GMRES-DR is that eigenvectors are calculated simultaneously while solving
the linear system and no separate calculation is needed. The eigenvectors calculated are 
approximate and their accuracy increases with each restart. In addition, eigenvectors 
computed with GMRES-DR could be used to accelerate the convergence for subsequent right-hand 
sides. This is a common situation in lattice QCD calculations where one needs to find
the quark propagator from all lattice sites using noise methods. The algorithm is called 
GMRES-Proj and is based on combining restarted GMRES with a projection over previously 
determined eigenvectors \cite{Morgan-Wilcox-gmres-proj}. For multiple right hand sides
the following two main steps are used:
\begin{itemize}
\item Solve the first right hand side using GMRES-DR.
\item For subsequent right-hand sides, solve by alternating between a minimal residual 
projection step over the right eigenvectors with smallest eigenvalues obtained from GMRES-DR
and one or more cycles of GMRES.
\end{itemize} 
Here, we extend this approach for multiple right-hand sides by replacing GMRES in the second step 
with BiCGStab. Since BiCGStab is a non-restarted method, the deflation step will be applied only 
once in the beginning. It will be used to obtain a better initial guess for the solution. In addition,
with right eigenvectors the Min.Res. projection does not do a good enough job of reducing the crucial
components in the direction of eigenvectors corresponding to small eigenvalues. So, instead, we will
project using both right and left eigenvectors.
In the case of Wilson fermions, 
left eigenvectors are obtained from right eigenvectors because the relation 
$\gamma_5 A_W \gamma_5 = A_W^\dagger$ is satisfied by the Wilson Dirac operator $A_W$.
\section{Algorithm for D-BiCGStab(k)}
Assuming that the first right-hand side was solved using GMRES-DR giving both the solution as
well as approximate right eigenvectors. 
The deflated BiCGStab for subsequent right-hand sides given in \cite{Morgan-Wilcox-gmres-proj}
is as follows:
\begin{itemize}
\item Consider the system $A x=b$. Let $x_0$ be an initial guess and $r_0=b-Ax_0$ be the initial residual.
\item Let $v_1,v_2,...,v_k$ an orthonormal basis for the set of $k$ right eigenvectors
and $u_1,u_2,...,u_k$ an orthonormal basis for the $k$ left eigenvectors. Define $U$ as $n \times k$ matrix
whose columns are the left eigenvectors $U=[u_1,u_2,...,u_k]$. Similarly, $V$ is $n \times k$ matrix whose
columns are the right eigenvectors $V=[v_1,v_2,...,v_k]$.
\item Solve the $k \times k$ linear system $U^\dagger A V y= U^\dagger r_0$ for $y$
and construct an improved initial guess $x_0^{new}=x_0 + V y$.
\item Apply BiCGStab to solve the system using $x_0^{new}$ as initial guess.  
\end{itemize}
\section{Results for D-BiCGStab(k)}
Deflated BiCGStab is first tested on quenched configurations generated using the Wilson plaquette action
at $\beta=6.0$ on $16^4$ and $20^3\times 32$ lattices (see \cite{lat07-Wilcox-plenary} for results on 
dynamical configurations). For Wilson fermions, we tune $\kappa$ to be close to
the critical value $\kappa_c$ in order to make it a difficult but physical problem. The value of $\kappa_c$ was
determined on each configuration from the condition that the real part of the smallest eigenvalue of $A$ 
vanishes. In the following we solve the even-odd preconditioned system.The first right-hand side is solved 
using GMRES-DR(m,k) where $k$ is the number of deflated eigenvectors and $m$ is the maximal dimension of the  
subspace. Both $m$ and $k$ are varied but the difference $m-k=20$ is kept fixed. The accuracy of the eigenvectors 
generated with GMRES-DR is increased by reducing the value of the residual norm at convergence when solving 
the first right-hand side. In the following, the first right hand side was solved with ratio of the norm of 
the residual at convergence to the norm of the initial residual $10^{-8}$, $10^{-10}$ and $10^{-14}$. 
This will be denoted by $L1$, $L2$ and $L3$ respectively with $L3$ corresponding to most accurate eigenvectors. 
For the second right-hand side, the relative residual norm at convergence is required to be
$10^{-8}$. For comparison, the second right-hand side is solved using standard BiCGStab and GMRES(20)-Proj(k). 
When convergence to the desired relative residual norm is not reached, this is indicated by the 
letter "F" in tables. Results are shown in Tables \ref{16to4table},\ref{20times32table} for a sample of three
configurations. 
{\small
\begin{table}[ht]
\begin{center}
\begin{tabular}{|c|c|c|c|ccc|c|}
\hline
C\#& $\kappa_c$&$m,k$ &  GMRES-DR(m,k) &  \multicolumn{3}{|c|}{D-BiCGStab(k)} & BiCGStab \\
\cline{5-7}
       &           &      & $1^{st}$rhs    &    L1         &    L2         &    L3&       \\
\hline
1      & 0.158383  & 30,10 & 650           &  746          &  708          &    490        & 720 \\
       &           & 40,20 & 560           &  726          &  550          &    472        &     \\
       &           & 50,30 & 550           &  764          &  590          &    402        &     \\
       &           & 60,40 & 540           &  692          &  504          &    356        &     \\
\hline
2      & 0.158399  & 30,10 & 710           &  818          &  718          &    644        & 806 \\
       &           & 40,20 & 620           &  832          &  584          &    486        &     \\
       &           & 50,30 & 610           &  796          &  542          &    370        &     \\
       &           & 60,40 & 600           &  774          &  458          &    418        &     \\
\hline
3      &0.157924   & 30,10 & 710           &  424          &  420          &    412        & 776 \\
       &           & 40,20 & 620           &  382          &  388          &    308        &     \\
       &           & 50,30 & 610           &  500          &  356          &    306        &     \\
       &           & 60,40 & 600           &  398          &  374          &    270        &     \\
\hline
\end{tabular}
\caption{Results for the $16^4$ lattice. Matrix-vector products listed correspond to relative residual norm $10^{-8}$.}
\label{16to4table}
\end{center}
\end{table}

\begin{table}[ht]
\begin{center}
\begin{tabular}{|c|c|c|c|ccc|c|}
\hline
C\# & $\kappa_c$ & $m,k$ & GMRES-DR(m,k) & \multicolumn{3}{|c|}{D-BiCGStab(k)} & BiCGStab \\
\cline{5-7}
    &            &       & $1^{st}$rhs   &    L1         &    L2         & L3  &          \\
\hline
1   & 0.157200   & 30,10 &  2310     & 1974 & 1738 & 1244 & 2332  \\
    &         & 40,20 &  1980     & 2396 & 1532 & 1170 &          \\
    &         & 50,30 &  1590     & 988  & 788  & 830  &          \\
    &         & 60,40 &  1540     & 1414 & 844  & 646  &          \\
    &         & 70,50 &  1510     & 978  & 842  & 676  &          \\
\hline
2   &0.157044& 30,10 & F & 1690 & 1690 & 1690 & 1812  \\
    &        & 40,20 & F & 1700 & 1700 & 1700 &       \\
    &         & 50,30 & 1410 &794 & 812 & 694  &      \\
    &         & 60,40 & 1360 &878 & 720 & 584  &      \\
    &         & 70,50 & 1330 &850 & 748 & 282  &      \\
\hline
3   &0.157095 & 30,10 & 1710 & 1524 &1214 & 1136 & 1534 \\
    &         & 40,20 & 1220 & 1188 &1134 & 912  &      \\
    &         & 50,30 & 1170 & 1172 &940  & 918  &      \\
    &         & 60,40 & 1120 & 1150 &912  & 808  &      \\
    &         & 70,50 & 1110 & 1214 &842  & 656  &      \\
\hline
\end{tabular}
\caption{Results for the $20^3\times 32$ lattice. Matrix-vector products listed correspond to
         relative residual norm $10^{-8}$.}
\label{20times32table}
\end{center}
\end{table}
}

\begin{figure}[htb]
\begin{center}
\vspace{0.2in}
\includegraphics[width=.6\textwidth]{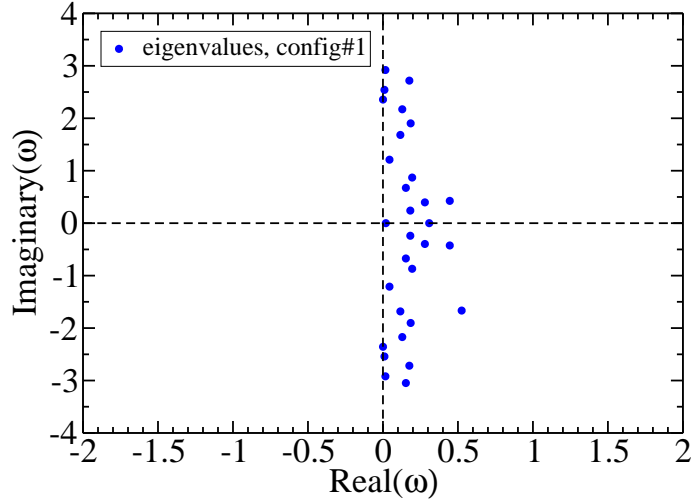}
\caption{Small eigenvalues near the origin for the first configuration of the $20^3\times 32$ lattice.}
\label{spectrum-20-32}
\end{center}
\end{figure}

\begin{figure}[htb]
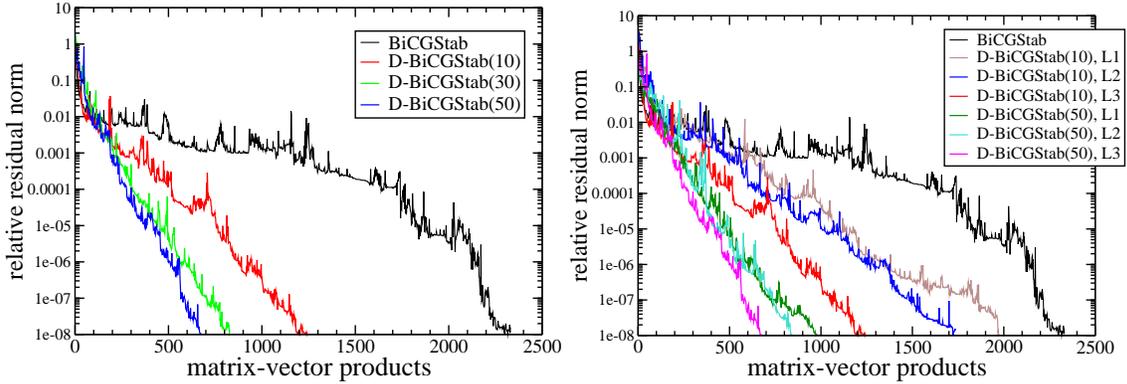

\begin{center}
\vspace{0.2in}
\includegraphics[width=.49\textwidth]{dbicg-vs-k-20-32.eps}
\includegraphics[width=.49\textwidth]{dbicg-vs-precision-20-32.eps}
\caption{Results for quenched Wilson configuration on $20^3\times32$. Left: effect of increasing the number of deflated
eigenvectors with L3 accuracy. Right: Effect of increasing the accuracy of the deflated eigenvectors.}
\label{dbicg}
\end{center}
\end{figure}

\begin{figure}[htb]
\vspace{0.5cm}
\hspace{1in}
\includegraphics[width=.6\textwidth]{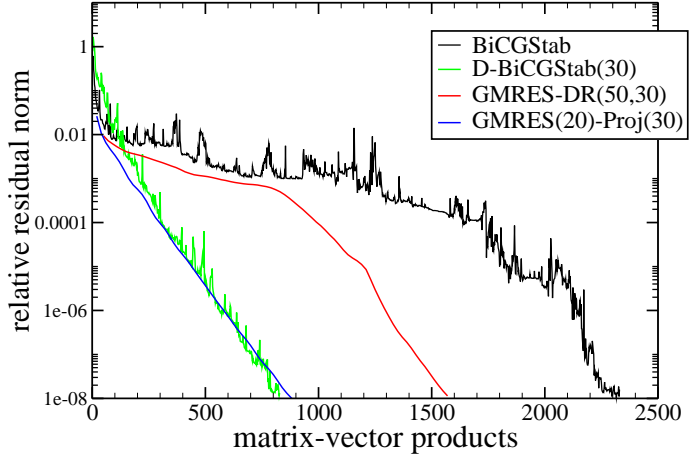}
\caption{Comparing results for the first configuration of the $20^3\times32$ lattice}
\label{compare-20-32}
\end{figure}

Comparing the results for D-BiCGStab(k) to BiCGStab, we find that the deflation step leads to a considerable improvement and
occasionally to a "breakthrough" as with the second configuration of the $20^3\times32$ lattice (Table \ref{20times32table}
with $(m,k)=(70,50)$ and accurate eigenvectors). For a fixed number of deflated eigenvectors, it is found that the improvement increases as the accuracy of the eigenvectors increases. So, the least number of matrix-vector products will correspond to the $L3$ columns. 
As expected also, the more eigenvectors we deflate the smaller the 
number of matrix-vector products. However, after certain optimal number of eigenvectors, an increase in the number of the deflated eigenvectors does not lead to large improvement. This optimal number of eigenvectors was found to increase as the volume of the
lattice increases but fortunately not linearly. Note also that when GMRES-DR does not converge, we don't get good eigenvectors and little
improvement is obtained by deflation (see results for the second configuration in Table \ref{20times32table}). 
For illustration, results for the first configuration of the $20^3\times32$ lattice
are shown in Figs. \ref{spectrum-20-32}-\ref{compare-20-32}. In Fig. \ref{spectrum-20-32}, a closer look at the eigenvalue spectrum
near the origin shows a very small eigenvalue as well as other small eigenvalues. 
In Fig. \ref{dbicg}, we show the effect of increasing the number of deflated eigenvectors as well as the effect of using more accurate eigenvectors. In Fig. \ref{compare-20-32}, we compare results for BiCGStab, GMRES-DR(50,30), GMRES(20)-Proj(30) 
and D-BiCGStab(30) using the most accurate vectors. Both deflated BiCGStab and GMRES-Proj improve substantially over BiCGStab.    

\section{A note on multi-mass solvers for Twisted-Mass QCD}
In Twisted Mass QCD, quarks are introduced as pairs with a modified mass
term \cite{Frezzotti}. The fermionic part of the action for a degenerate pair 
of quarks is given by (see \cite{Frezzotti} for unexplained notations):
\begin{equation}
S_f=\sum_x\sum_\nu \bar{\Psi}\{\frac{1}{2}[\gamma_\nu(\nabla_\nu+\nabla_\nu^*)-\nabla_\nu^*\nabla_\nu]
     +m+i\mu\gamma_5\tau_3\}\Psi,
\end{equation}
where $\bar{\Psi}=(\bar{u}~,~\bar{d})$ 
is a doublet of the up and down quarks and
$m$ and $\mu$ are the standard and twisted mass terms respectively. 
Because of the twisted mass term, it is not possible to apply multi-mass solvers
to twisted mass problems simultaneously with even-odd preconditioning.
Multi-mass solvers as CGS can be used if even-odd preconditioning is not
implemented \cite{tm-multimass}. In the following we describe how one can accelerate the convergence
of twisted-mass problems with multiple masses and even-odd preconditioning.
The method is based on solving the systems serially but using an improved initial
guess by making a minimal residual projection over available solutions
of the previous systems. It is described as follows:
\begin{itemize}
\item Consider the systems $A_i x_i = b$ where $A_i$ is the even-odd preconditioned 
Twisted-Mass Dirac operator that corresponds to $\kappa^c_i,\mu_i$. We assume the system
at maximal twist, so $\kappa^c_i$ is the critcal $\kappa$ value that corresponds to the twisted
mass value $\mu_i$.
\item Assume that we solved the systems for $i=1,..k$. In order to accelerate the solution of the
system $k+1$, we perform a minimal residual projection over the $k$ previous solutions as follows:
     \begin{itemize}
     \item let $x^0_{k+1}$ be the initial guess and $r^0_{k+1}=b-A_{k+1}x^0_{k+1}$ the initial resiual.
     Define $Q$ as the $n \times k$ matrix whose columns are the previous $k$ solutions, i.e.
           $Q=[x_1,x_2,...,x_k]$. 
     \item Solve the $k \times k$ system $Q^\dagger A_{k+1}^\dagger A_{k+1} Q  d = Q^\dagger A_{k+1}^\dagger r^0_{k+1}$ 
           for $d$ and construct an improved initial guess $\tilde{x}^0_{k+1}=x^0_{k+1}+Qd$.
     \end{itemize}
\end{itemize}
The method was tested on $20^3\times32$ lattice with quenched configurations at
$\beta=6.0$ with the Wilson plaquette action. The twisted mass fermion action
for a degenerate doublet of quarks at maximal twist for 11 values of $\mu$ is used.
The values of $\kappa^c$ for each value of $\mu$ is determined using a
linear fit of the four $(\kappa^c,\mu)$ pairs in \cite{parity-maximal-twist}. Three parameters affect the 
performance of the method. First is the separation between successive masses, second the number of available 
solutions to project over and third whether we solve the heaviest or the
lightest mass first. In Table \ref{serial-multimass}, we compare the number of matrix-vector products
when zero initial guess is used to the case where projection over previous solutions is done
for a typical mass separation. Although the projection step is done only once per shift, we found a considerable reduction in the total number of matrix-vector products. The overall reduction seems to be similar whether we solve the heaviest or 
the lightest mass first for that mass separation.


\begin{table}[ht]
\begin{center}
\begin{tabular}{|c|c|c|c|c|c|}
\hline
Mass Number & $\kappa^c$ & $\mu$ & $x^0=0$   & With projection  & With projection \\
            &            &       &           & high $\rightarrow$ low& low$\rightarrow$high \\
\hline
1  & 0.157290 & 0.005 & 1270 & 1000 & 1270 \\
2  & 0.157210 & 0.009 & 1150 & 730  & 1030 \\
3  & 0.157130 & 0.013 & 1030 & 550  & 880  \\
4  & 0.157050 & 0.017 & 880  & 430  & 700  \\
5  & 0.156970 & 0.021 & 790  & 400  & 520  \\
6  & 0.156890 & 0.025 & 730  & 280  & 400  \\
7  & 0.156810 & 0.029 & 640  & 280  & 310  \\
8  & 0.156730 & 0.033 & 580  & 310  & 220  \\
9  & 0.156650 & 0.037 & 520  & 340  & 190  \\
10 & 0.156570 & 0.041 & 490  & 400  & 160  \\
11 & 0.156490 & 0.045 & 460  & 460  & 190  \\
\hline
Total MVP &  &  & 8,540 & 5,180 & 5,870  \\
\hline
\end{tabular}
\caption{Effect of the projection step with GMRES-DR(40,10) on $20^3\times 32$ lattice
starting from highest to lowest mass or vice versa with $\Delta\mu=0.004$.}
\label{serial-multimass}
\end{center}
\end{table}

\section{Conclusions}
For problems with multiple right-hand sides, a combination
of GMRES-DR for the first right-hand side and a deflated BICGStab for
subsequent right-hand sides was tested on typical lattice volumes. It was
found to give a considerable reduction of the matrix-vector products by a factor
of approximately 5 for the L3 case. The improvement level increases as 
the accuracy of the eigenvectors increase. Deflated BiCGStab was tested with left-right projection. 
The method does not add extra work for the case of Wilson fermions where the left eigenvectors are
related to the right eigenvectors through $\gamma_5$ multiplication. 
For Twisted-Mass problems with multiple shifts and even-odd preconditioning, it was found that improving the
initial guess using a minimal residual projection over previous solutions reduces the matrix-vector products
by a factor of $20\%-50\%$.

\acknowledgments
WW gratefully acknowledges summer sabbatical support from Baylor University. Calculations were done using the Mercury cluster at the National Center for Supercomputing Applications (NCSA) and the Baylor University High Performance Cluster.

\end{document}